\documentclass{elsart1p}

\usepackage{amssymb,amsmath}
\usepackage{graphicx}
\usepackage[frenchb,english]{babel}
\usepackage{enumerate}
\usepackage{natbib}

\usepackage{ifpdf}
\ifpdf
\usepackage[%
  pdftitle={Vlasov limit and discreteness effects in cosmological N-body simulations},%
  pdfauthor={Bruno Marcos},%
  pdfsubject={Vlasovia proceedings},%
  pdfkeywords={N-body simulations, fluid theory, linearized gravity, discreteness effects},%
  pdfstartview=FitH,%
  bookmarks=true,%
  bookmarksopen=true,%
  breaklinks=true,%
  colorlinks=true,%
  linkcolor=blue,anchorcolor=blue,%
  citecolor=blue,filecolor=blue,%
  menucolor=blue,pagecolor=blue,%
  urlcolor=blue]{hyperref}
\else
\usepackage[%
  breaklinks=true,%
  colorlinks=true,%
  linkcolor=blue,anchorcolor=blue,%
  citecolor=blue,filecolor=blue,%
  menucolor=blue,pagecolor=blue,%
  urlcolor=blue]{hyperref}
\fi

\makeatletter
\def\elsartstyle{%
    \def\normalsize{\@setfontsize\normalsize\@xiipt{14.5}}
    \def\small{\@setfontsize\small\@xipt{13.6}}
    \let\footnotesize=\small
    \def\large{\@setfontsize\large\@xivpt{18}}
    \def\Large{\@setfontsize\Large\@xviipt{22}}
    \skip\@mpfootins = 18\p@ \@plus 2\p@
    \normalsize
}
\@ifundefined{square}{}{}
\makeatother

\def\ep{\epsilon}

\def\om{\omega}

\def\be{\begin{equation}} 
\def\ee{\end{equation}} 
\def\bea{\begin{eqnarray}} 
\def\eea{\end{eqnarray}}  
\def\bean{\begin{eqnarray*}} 
\def\eean{\end{eqnarray*}} 
\def\dd{\partial}

\def\bk{{\bf k}}  
\def\bx{{\bf x}}  
\def\bv{{\bf v}}
\def\br{{\bf r}}

\def\bu{{\bf u}}
\def\bR{{\bf R}}
\def\mD{{\cal D}}

\def\tmD{{\tilde\mD}}

\def\na{{\nabla}}

\def\bq{{\mathbf q}}
\def\bse{\begin{subequations}}
\def\ese{\end{subequations}}

\def\da{{\dot a}}
\def\dda{{\ddot a}}

\def\bq{\mathbf{q}}

\def\lsim{\raise 0.4ex\hbox{$<$}\kern -0.8em\lower 0.62ex\hbox{$\sim$}} 
\def\gsim{\raise 0.4ex\hbox{$>$}\kern -0.7em\lower 0.62ex\hbox{$\sim$}}

\def\bk{\mathbf{k}}

\def\f0N{f_0^{(N)}}
\def\bec{\begin{center}}
\def\eec{\end{center}}

\def\piG{{\pi G}}

\newcommand{\ve}[1]{\mathbf{#1}}

\begin{document}

\begin{frontmatter}
\title{Vlasov limit and discreteness effects in cosmological N-body simulations}

\author{Bruno Marcos}
\address{Dipartimento di Fisica, Universit\`a ``La Sapienza'',
P.le A. Moro 2,
I-00185 Rome,
Italy,\\
\& ISC-CNR,
Via dei Taurini 19,
I-00185 Rome,
Italy.}

\ead{Bruno.Marcos@roma1.infn.it}

\begin{abstract}
We present the problematic of controlling the discreteness effects in
cosmological N-body simulations. We describe a perturbative treatment which
gives an approximation describing the evolution under self-gravity of a
lattice perturbed from its equilibrium, which allows to trace the evolution of
the fully discrete distribution until the time when particles approach one
another ("shell-crossing"). Perturbed lattices are typical initial conditions
for cosmological N-body simulations and thus we can describe precisely the
early time evolution of these simulations. A quantitative comparison with
fluid Lagrangian theory permits to study discreteness effects in the linear
regime of the simulations. We show finally some work in progress about
quantifying discreteness effects in the non-perturbative (highly non-linear)
regime of cosmological N-body simulations by evolving different
discretizations of the same continuous density field.
\end{abstract}

\end{frontmatter}


\section{Introduction}
In this proceedings we present some specific work about the material presented
in the proceedings of Michael Joyce\footnote{See {\tt arXiv:0805.1453}.}, where a detailed introduction can be
found. In what follows we remind very briefly the motivations of our work.

At the scales of interest in cosmology, the evolution of the dark mater can be
very well described as a fluid, which obeys the Vlasov equation coupled with
the Poisson equation (e.g. \cite{peebles_80}):
\bse
\label{vlasov-poisson}
\begin{align}
\label{vlasov}
&\frac{\dd f(\br,\bv,t)}{\dd t}+\bv\cdot\nabla
f(\br,\bv,t)-\nabla\Phi\cdot \frac{\dd f(\br,\bv,t)}{\dd \bv}=0\\
\label{poisson}
&\nabla^2\Phi(\br,t)=4\piG\rho(\br,t),
\end{align}
\ese
 where $f(\br,\bv,t)$ is the one-particle phase space density in
 physical coordinates (i.e. it gives the probability density to find a
 particle at $\br$ with velocity $\bv$ at time $t$), $\rho(\br,t)$ the
 local density of dark matter and $G$ the gravitational constant. The
 solution of the Vlasov equation is generally ``estimated'' performing
 N-body simulations, using a number of particles much smaller than the
 number of dark matter particles (it can be seen as a discretization
 of the system). This method leads inevitably to discreteness
 effects, which are still not well understood.

Up to now the primary approach to the study of discreteness in N-body
simulations has been through numerical studies of convergence (see
e.g. \cite{power_03,diemand_04}), i.e., one changes the number of
particles in a simulation and studies the stability of the measured
quantities. Where results seem fairly stable, they are assumed to have
converged to the continuum limit.  While this is a coherent approach,
it is far from conclusive as, beyond the range of perturbation theory,
we have no theoretical ``benchmarks'' to compare with. Nor is there
any systematic theory of discreteness effects, e.g., we have no
theoretical knowledge of the $N$ dependence of the convergence. It can
be found in the literature several studies using this approach of the
most obvious effect of discreteness --- and certainly the one most
emphasized in the literature --- which is the two body collisionality: pairs of
particles can have strong interactions with one another, which is an
effect absent in the collisionless limit. For analysis and discussion
of these effects see, e.g.,
\cite{melott_93,melott_97,splinter_98,baertschiger_02t,binney_02}.

In this proceedings we present a work which permits to understand
systematically the discreteness in the {\em linear regime} of
gravitational clustering. We do so by developing a perturbative
solution to the fully discrete cosmological N-body problem. This
essentially analytic solution can be compared to the analogous fluid
($N\to\infty$) theory, of which analytic solution is well-known. We
will remind first the standard {\em perturbative fluid theory} (FLT),
we will construct the {\em particle linear theory} (PLT) and by
comparing both we will quantify the discreteness effects in the linear
regime. We will also present a test which permits to quantify the
discreteness effects in the non-perturbative regime of gravitational
clustering. The material presented in these proceedings is based on
work which can be found in \cite{jmgbsl_05,marcos_06,baertschiger_06}.

\section{Fluid theory}
\label{FLT}
 It is convenient, to simplify the problem Vlasov equation, to
construct a set of {\em fluid equations}, which give a less detailed
(but sufficiently accurate) description of the system. We define the {\em
mass density} and the {\em mean fluid velocity} from the velocity
moments of $f(\br,\bv,t)$:
\be
\label{momenta-f}
\rho(\br,t)=m\int d\bv \, f(\br,\bv,t),\qquad\qquad
\rho \overline{\bv}(\br,t)=m\int d\bv\,  \bv\,
f(\br,\bv,t).
\ee
where $m$ is the mass of the dark matter particles. Inserting
Eqs.~\eqref{momenta-f} in the Vlasov-Poisson equation
\eqref{vlasov-poisson}, we obtain the {\em continuity} equation and
the {\em Euler} equation in the variables $\rho(\br,t)$ and
$\overline{\bv}(\br,t)$. These are the fluid equations in the {\em Eulerian}
formalism. In cosmology, it is common to use the {\em Lagrangian}
formalism which is known to give better results in a perturbative
approach (see e.g. \cite{sahni_95}).The idea of the Lagrangian
formulation is to calculate the trajectories of infinitesimal fluid
elements. The velocity is given by the velocity of these fluid
elements and the density varies according with the convergence or the
divergence of the fluid elements to each point.

We define the
Lagrangian coordinate $\bq$ as the position of the fluid element at
the initial time\footnote{It can be viewed just as a ``label'' of the
particle.}. Because we work in an expanding universe, it is natural to write the physical position of a fluid element $\br$ in function of its Lagrangian coordinate $\bq$ as:
\be
\label{mapping}
\br(t)=a(t)(\bq+\bu(\bq,t)), 
\ee 
where $a(t)$ is the scale factor (which is given by the particular
cosmological model considered) and $\bu(\bq,t)$ is a ``displacement
field'' of the fluid elements. The displacement field $\bu(\bq,t)$ in
Eq.~\eqref{mapping} can be understood as a perturbation of the homogeneous model which expands with the scale factor $a(t)$. Using the
coordinate transformation \eqref{mapping} we obtain a set of fluid
equations for the displacement field $\bu(\bq,t)$. Linearizing them
about the displacements $\bu(\bq,t)$ and neglecting pressure
corrections (which is a very good approximation for sufficiently large
scales), one obtains a simple system of equations (see
e.g. \cite{buchert_92} for details). To look for a solution of these
equations, it is convenient to divide the displacements into a
curl-free part, $\bu_\parallel$ and a divergence-free part,
$\bu_\perp$, i.e., $\bu=\bu_\parallel+\bu_\perp$ (i.e. $\na
\times \bu_\parallel=\mathbf 0$ and $\nabla\cdot \bu_\perp=0$). Then we
obtain the set of equations  (choosing appropriate boundary conditions) \cite{buchert_92}:
\be
\label{lagb-expl}
\ddot\bu_\perp+2\frac{\da}{a}\dot\bu_\perp=\mathbf 0,\qquad\qquad \ddot\bu_\parallel+2\frac{\da}{a}\dot\bu_\parallel+3\frac{\dda}{a}\bu_\parallel=\mathbf 0.
\ee
Given initial conditions, it is simple to compute the
solution for the displacement field with time. The linear
approximation (and more generally a perturbative approach of the
Lagrangian fluid equations) breaks down when the volume of a fluid
element becomes zero. This is called {\em shell crossing}.

\section{Linear perturbative theory}
\label{PLT}
In this section we will present a linearized theory of the N-body
problem (PLT), which is the discrete analogue of the linear fluid
theory presented in the previous section. The starting point is the full
evolution of the N-body system in comoving coordinates:
\be
\label{full-ev}
\ddot \bx_i+2\frac{\da}{a}\dot \bx_i=-\frac{1}{a^3}\sum_{i\ne j} \frac{G m (\bx_i-\bx_j)}{|\bx_i-\bx_j|^3},
\ee
where $\bx_i$ is the position of the $i$-th particle and $a(t)$ the scale factor.

A standard way to generate initial conditions for N-body simulations
consists in perturbing a lattice (see e.g. \cite{bertschinger_98}). It
is therefore natural to build a perturbative theory where the
perturbed variable is the displacement of each particle about the
lattice, which is an equilibrium position. We will therefore have an
accurate description of the clustering when the displacements (or, in
fact, the relative displacements) are smaller than the inter-particle
distance. When the relative displacements become larger than the
inter-particle distance, the approximation breaks down, which is
equivalent to the ``shell-crossing'' in fluid theory.

We define the displacement $\bu(\bR,t)$ about the lattice position
$\bR$ as $\bx_i(t)=\bR+\bu(\bR,t)$, where $\bR$ is the lattice site of
particle $i$. Using this notation we can linearize the full evolution
\eqref{full-ev} as
\be
\label{linear-ev}
{\bf {\ddot u}}({\bf R},t) 
+2 \frac{\da}{a} {\bf {\dot u}}({\bf R},t) 
= -\frac{1}{a^3} \sum_{{\bf R}'} 
{\cal D} ({\bf R}- {\bf R}') {\bf u}({\bf R}',t),
\ee
where ${\cal D}(\bR)$ is called the {\em dynamical matrix}, which is ${\cal D}_{\mu\nu}(\bR=\mathbf 0)=-\sum_{\bR\ne\mathbf 0} {\cal D}_{\mu\nu}(\bR)$ and ${\cal D}_{\mu\nu}(\bR\ne\mathbf 0)=G m\left(\frac{\delta_{\mu \nu}}{R^3}-3\frac{R_{\mu}R_{\nu}}{R^5}\right)$.
Taking periodic boundary conditions (as in the N-body simulations), it
is possible to diagonalize Eq.~\eqref{linear-ev} simply by taking its
Fourier transform, defined as
$\tilde\bu(\bk,t)=\sum_{\bR}\bu(\bR,t)e^{-i\bk\cdot\bR}$ for the
displacement field, and in an analogue manner for the dynamical matrix
$ \cal \tilde D(\bk)$. Using Eqs.~\eqref{linear-ev} and the definition of the Fourier transform, we obtain for each $\bk$: \be
\label{modes-eq}
{\bf \ddot{{\tilde u}}} ({\bf k},t) 
+ 2 \frac{\da}{a} {\bf \dot{{\tilde u}}} ({\bf k},t) 
= -\frac{1}{a^3} {\cal {\tilde D}} ({\bf k}) {{\bf {\tilde u}}}({\bf k},t).
\ee
It is possible to solve Eq.~\eqref{modes-eq} by diagonalizing the
$3\times3$ matrices $\tmD(\bk)$. For each $\bk$, this determines three
orthonormal eigenvectors $\mathbf{\tilde e}_n(\bk)$ with three
associated real eigenvalues\footnote{The eigenvalues are real and the
eigenvectors orthonormal because $\cal\tilde D(\bk)$ is a real and
symmetric matrix.}  $\om_n^2(\bk)$ ($n=1,2,3$), satisfying the
eigenvalue equation $\tmD(\bk)\mathbf{\tilde e}_n(\bk)=\om_n^2(\bk)
\mathbf{\tilde e}_n(\bk)$. Once $\om_n^2(\bk)$ is calculated, the
evolution of the displacement of the particles is straightforward to
compute.
\begin{figure}
\rotatebox{0}{\includegraphics[width=0.5\textwidth]{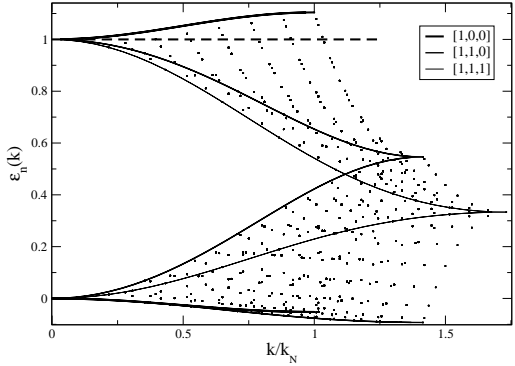}}
\caption{Spectrum of eigenvalues for simple cubic lattice with $16^3$
particles. The lines correspond to chosen directions in $k$ space. 
\label{fig1}}
\end{figure} 
\subsection{Spectrum of eigenvalues of $\tmD(\bk)$}

To understand the dynamics of the PLT is important to study the
spectrum of eigenvalues of the matrix $\tmD(\bk)$. We will describe in
what follows this spectrum for a simple cubic (hereafter sc) lattice,
which is very widely used in N-body simulations of structure
formation in cosmology.

In Fig.~\ref{fig1} we plot the spectrum of a sc lattice, for $N=16^3$
particles.  We show the normalized eigenvalues
$\mathbf{\varepsilon}_n(\bk)=\om_n^2(\bk)/4\pi G\rho_0$ as a function
of the modulus of the $\bk$ vectors, normalized to the Nyquist
frequency $k_N=\pi/\ell$ ($\ell$ is the inter-particle distance).
With this normalization the spectrum remains substantially the same as
we increase the number of particles: the only change is that the
eigenvalues become denser in the plot, filling out the approximate
functional behaviors with more points. We see how for each vector
$\bk$ there are three eigenvalues $\om_n^2(\bk)$, $n=1,2,3$.  Each
family of eigenvalues (i.e. with same $n$) defines a surface,
corresponding to the three branches of the frequency-wavevector
dispersion relation. Sections of these surfaces are plotted for some
chosen directions of the vector $\bk$ in Fig.~\ref{fig1}.

The fluid limit of this system is given by taking the $\bk\to\ve 0$ limit
keeping the inter-particle distance $\ell$ constant: a plane wave fluctuation
$e^{i\bk\cdot\br}$ with $\bk\ll 1/\ell$ has a variation scale much larger than
the inter-particle distance, and therefore does not ``see'' the particles. It
is simple to show (e.g. \cite{marcos_06,pines_63}) that in this limit the
eigenvectors and eigenvalues are: {\em (i)} one {\em longitudinal} eigenvector
polarized parallel to $\bk$ with normalized eigenvalue
$\varepsilon_1(\bk\to\ve 0)=1$ and {\em (ii)} two {\em transverse}
eigenvectors polarized in the plane transverse to $\bk$ with normalized
eigenvalues $\varepsilon_{2,3}(\bk\to\ve 0)=0$.  It is interesting to note
that in this limit, using Eq.~\eqref{linear-ev}, we obtain exactly the two
fluid equations \eqref{lagb-expl}, particularized to an Einstein-de Sitter
(EdS) universe and with the same boundary conditions. This means that the
N-body system is a specific discretization --- at least in the linear
regime--- of the underlying fluid theory.
\begin{figure}
\includegraphics[width=0.6\textwidth]{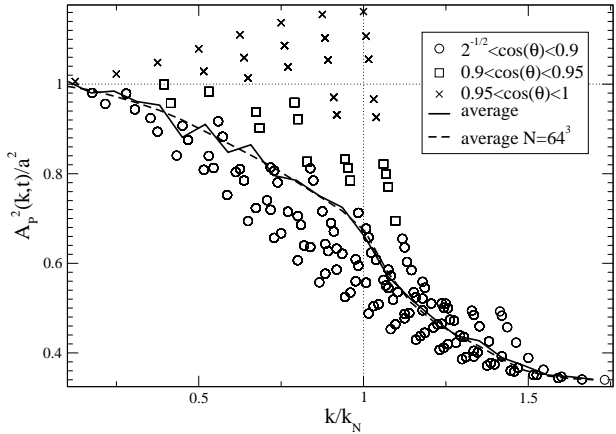}
\caption{Amplification function $A^2(\bk,t)$ divided by the fluid
  amplification factor at $a=5$, for a sc lattice, in function of the
  modulus of the wave-vectors normalized to the Nyquist frequency.
\label{fig2}}
\end{figure}

\subsection{Discreteness effects}
 It is now straightforward to quantify precisely the discreteness effects ---
in the linear regime --- by comparing the FLT explained in section \ref{FLT}
and the PLT derived in section \ref{PLT}. We will focus here on the
discreteness effects in the power spectrum (hereafter PS, for its definition see
e.g. \cite{peebles_80,GSJP_05}) but an analogous analysis can be performed for
any statistical quantity or, for example, the trajectories of a single
particle (see \cite{marcos_06}).

It is possible to show, considering an initial power spectrum $P_0(k)$
at $t=t_0$, that we have, as a very good approximation (for not too
large displacements compared to the lattice spacing), $P(\bk,t)\simeq
A_P^2(\bk,t)P_0(k)$, where $A_P^2(\bk,t)$ is a function which depends
on the particular lattice and the cosmological model considered. One
of the most important result obtained in this work is that
discreteness effects {\em increase with time}. In the literature the
time was not considered as a parameter in which depends the
discretness effects, but only ``static'' parameters as the number
density of particles.

 It is possible to check that,
in the fluid limit, we recover the well-known amplification of the PS in the
fluid theory, i.e., $\lim_{\bk\to\mathbf
0}A_P(\bk,t)=\left(t/t_0\right)^{2/3}=a$.
We plot the function $A_P(\bk,t)$, normalized to the fluid amplification, in
Fig.~\ref{fig2}. We have chosen a value of $a=5$ for the scale factor.  This
is a typical scale factor at which shell crossing occurs in cosmological
simulations. Notice the similarity of this figure with the optical branch in
Fig.~\ref{fig1}: the evolution ``deforms'' the spectrum of eigenvalues.  Note
how the eigenvalues with $\ep>1$ give rise to $A_{P}(\bk,t)>a$ for these
modes, i.e., there are modes which grow faster than the fluid limit. In the
figure, we have classified the modes as a function of the angle subtended by
their wave vector $\bk$ with the lattice axis that form the minimal angle with
it. We see that there is a strong dependence of the value of the eigenvalue on
this angle: the closer $\bk$ is to parallel to one of the axes, the larger is
the eigenvalue of the mode, on average. This is a manifestation of the
breaking of isotropy introduced by the N-body discretization on the
lattice. Even if there are some modes that grow faster than the fluid,
averaging over bins with similar $|\bk|$ the resultant growth is slower ---
because we consider sufficiently early times --- than the fluid limit. Note
that this averaging is generally performed when computing statistical
properties of the particle distribution, such as the PS, for example.
\begin{figure*}
\includegraphics[width=0.32\textwidth]{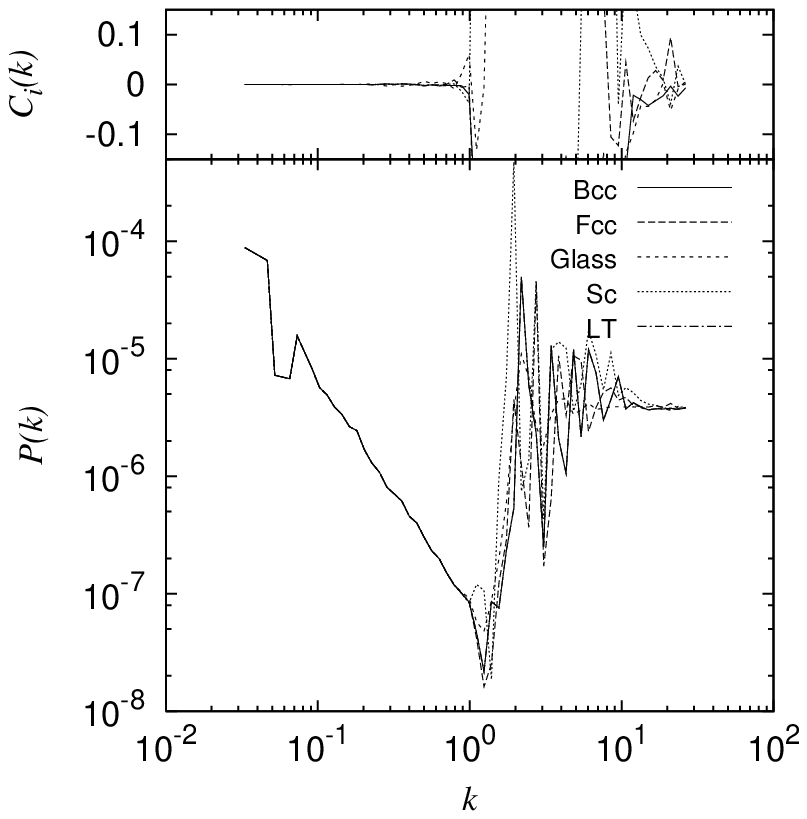}
\includegraphics[width=0.32\textwidth]{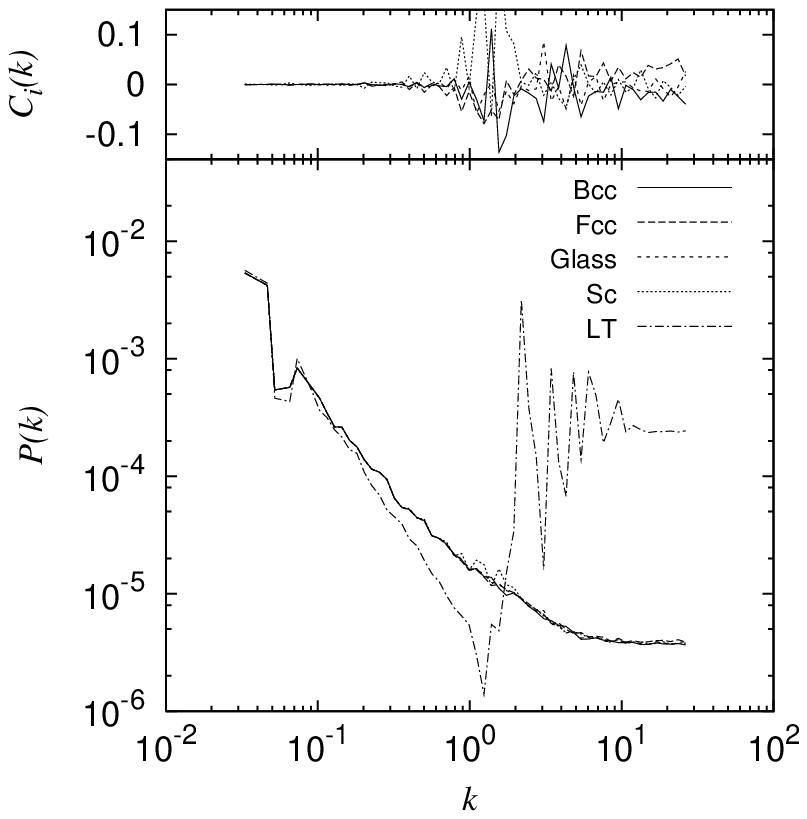} 
\includegraphics[width=0.32\textwidth]{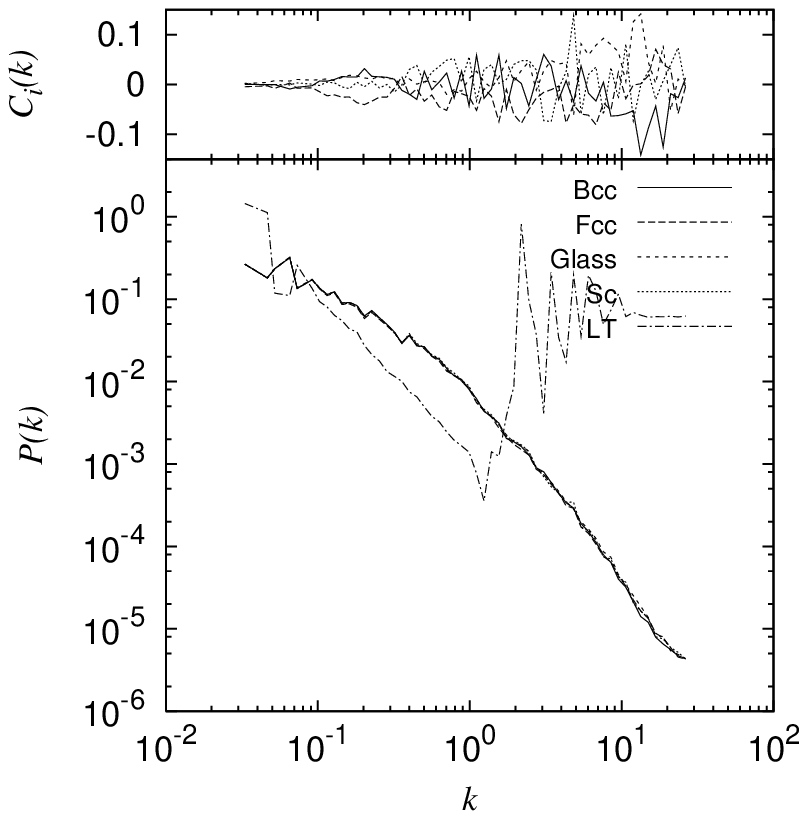}
\caption{Evolution of the PS and $C_i(k)$ for, from up to bottom and
left to right, $a=1$, $a=2^3$ and $a=2^7$. We have also
included the evolution of the PS predicted by linear theory (``LT'').
 \label{corr}}
\end{figure*}

\section{Discreteness effects in the non-perturbative regime}
It is much more difficult to quantify the discreteness effects in the non
perturbative regime because there is no known solution of the fluid theory. In
this section we present briefly some work in progress which goal is to quantify
the discreteness effects in this regime. We run a set of simulations with
exactly the same parameters (time-step, force accuracy, etc.) for an initial
PS $P(k,0)\propto 1/k^2$ (which corresponds to the kind of density
fluctuations of the standard cosmological model) in an expanding EdS universe,
taking different perturbed lattices to generate the initial conditions: body
cubic centered (bcc), face cubic centered (fcc), sc and ``glass''\footnote{It
is a highly ordered stochastic distribution, see \cite{white_94}.}. It can be
considered as different discretizations of the same continuous density field.

In Fig.~\ref{corr} we show the evolution of the PS for times
corresponding to $a=1$, $a=2^3$, $a=2^7$. In the largest figures are
shown the PS and in the smallest ones the relative residuals $C_i(k)$
between the PS of a particular lattice $i$ ($i=$ bcc, fcc, glass or
sc) and the average of all of them:
\be
\label{residuals}
C_i(k)=\frac{1}{m}\frac{m\,P_i(k)-\sum_{j=1}^m P_i(k)}{\sum_{j=1}^m P_j(k)}.
\ee
 At the initial time, the PS up to the Nyquist frequency is
 approximately the same for all the lattices. However, at scales
 larger than the Nyquist frequency, the PS corresponding to the
 different lattices is very different due to the great differences
 between the distributions at small scales. Time evolution ($a=2^3$)
 erases almost all the differences at small scales. At all scales, the
 evolution of the PS of the different initial distributions is very
 similar. The residuals show that {\em i)} the discreteness effects
 have been globally reduced, {\em ii)} the differences can be up to
 $10\%$ and {\em iii)} the scale at which these differences appear is
 much larger than the initial discreteness scale $k_N$: there is a
 propagation of discreteness from small scales to large scales. In the
 figure we have shown also the evolution of the initial PS using fluid
 linear theory.  At $a=2^7$ there are no linear modes left in the box
 (it can be seen by the fact that linear fluid theory does not
 reproduce the evolution of any smallest Fourier mode). The
 differences between the PS of the different initial lattices are
 larger than in the previous time slice $a=2^3$ (in which there were
 still some linear modes). This is related to the fact that all the
 system is highly non linear. Finally, it is interesting to note that, as in
 the linear regime, the discreteness effects increase with time.


\section*{}
{\bf Acknowledgments}\\
  I am indebted to T. Baertschiger, A. Gabrielli, M. Joyce and
  F. Sylos Labini for our fruitful collaboration. I also thank them to
  have read and for comments on this manuscript. The simulations where
  performed on the cluster of the ``E. Fermi'' Center of Rome.



\end{document}